\documentclass[fleqn,usenatbib]{mnras}

\usepackage{mathptmx}

\usepackage[T1]{fontenc}

\DeclareRobustCommand{\VAN}[3]{#2}
\let\VANthebibliography\thebibliography
\def\thebibliography{\DeclareRobustCommand{\VAN}[3]{##3}\VANthebibliography}

\usepackage[pdftex]{graphicx,color}
\usepackage{amsmath}	
\usepackage{amssymb}	

\title[Centroid migration on an impacted slope]{Centroid migration on an impacted granular slope due to asymmetric ejecta deposition and landsliding}

\author[T. Omura et al.]{
Tomomi Omura,$^{1,2}$
Shinta Takizawa,$^{2}$
Hiroaki Katsuragi$^{2,3}$
\\
$^{1}$Institute of Education Center of Advanced Education, Osaka Sangyo University, 3-1-1 Nakagaito, Daito-shi, Osaka 574-8530, Japan\\
$^{2}$Department of Earth and Environmental Sciences, Nagoya University, Furocho, Chikusa, Nagoya 464-8601, Japan\\
$^{3}$Department of Earth and Space Science, Osaka University, 1-1 Machikaneyama, Toyonaka 560-0043, Japan
}

\date{Accepted XXX. Received YYY; in original form ZZZ}

\pubyear{2020}

\begin{document}
\label{firstpage}
\pagerange{\pageref{firstpage}--\pageref{lastpage}}
\maketitle

\begin{abstract}
  For a fundamental understanding of terrain relaxation occurring on sloped surfaces of terrestrial bodies, we analyze the crater shape produced by an impact on an inclined granular (dry-sand) layer. Owing to asymmetric ejecta deposition followed by landsliding, the slope of the impacted inclined surface can be relaxed. Using the experimental results of a solid projectile impact on an inclined dry-sand layer, we measure the distance of centroid migration induced by asymmetric cratering. We find that the centroid migration distance $x_\mathrm{mig}$ normalized to the crater minor-axis diameter $D_\mathrm{cy}$ can be expressed as a function of the initial inclination of the target $\tan\theta$, the effective friction coefficient $\mu$, and two parameters $K$ and $c$ that characterize the asymmetric ejecta deposition and oblique impact effect: $x_\mathrm{mig}/D_\mathrm{cy}=K \tan\theta/(1-(\tan\theta/\mu)^2)+c$, where $K=0.6$, $\mu=0.8$, and $c=-0.1$ to $0.3$. This result is consistent with a previous study that considered the effect of asymmetric ejecta deposition. The obtained results provide fundamental information for analyzing the degradation of sloped terrain on planetary surfaces, such as crater-shape degradation due to the accumulation of micro-impacts.
\end{abstract}

\begin{keywords}
planets and satellites: surfaces -- planets and satellites: terrestrial -- methods: miscellaneous planets
\end{keywords}



\section{Introduction}
Impact cratering is one of the most ubiquitous phenomena occurring in the solar system. From the shapes and size/spatial distributions of abundant craters on a surface, one can extract useful information about its conditions and history~\citep{Melosh:1989,Melosh:2011,Osinski:2013,Katsuragi:2016}. Thus far, various studies have focused on the dynamics concerning the impact crater formation process. For instance, to estimate the scale of impact (e.g., the impactor's kinetic energy, momentum, or their coupling parameter) from the crater dimensions and ejecta deposition, dimensionless scaling methods have been used~\citep{Holsapple:1993,Housen:2011}. The scaling relations were established on the basis of experimental results of vertical impacts onto horizontal surfaces. However, most natural impacts occur obliquely onto sloped terrains. Classically, ~\citet{Gault:1978} reported that impact obliquity affects the shape of craters. Recently, the effect of inclination of the target surface on the crater shape was studied~\citep{Hayashi:2017,Aschauer:2017}. In addition, the effects of both oblique impact and sloped terrain were systematically studied in a recent experiment~\citep{Takizawa:2020}. Moreover, an artificial impact experiment was performed on the surface of asteroid 162173 Ryugu~\citep{Arakawa:2020}. In such a natural impact experiment, various complex terrains must affect the crater formation dynamics. 

Instantaneous crater formation and resultant crater shape have been extensively studied, as mentioned above. However, the crater shape formed by an impact could be modified (degraded) after its formation event. Although degradation of crater shape usually proceeds very slowly, it has a long duration. As a consequence of degradation, crater shapes become vague and are eventually erased. While the timescale of crater degradation is substantially longer than its formation timescale, it must be sufficiently shorter than the lifetime of the surface of the target body. Otherwise, crater degradation could not be observed. In general, the timescale of crater degradation depends on the crater size, surface gravity, etc. Its typical value can exceed $10^6$~yr~\citep{Ross:1968}. Various mechanisms have been considered for fundamental processes of degradation, e.g., global seismic shaking~\citep{Richardson:2004,Richardson:2005}, viscous relaxation~\citep{Pathare:2005,Mohit:2007}, distal ejecta deposition~\citep{Minton:2019}, and crater-wall relaxation due to numerous small-scale impacts~\citep{Soderblom:1970}. The crater degradation processes relate to the equilibrium of size frequency distribution of craters~\citep{Gault:1970,Marcus:1970,Hirabayashi:2017,Minton:2019}. An equilibrium distribution is achieved when the crater production rate and its obliteration rate due to crater degradation are balanced. Recently, \citet{Minton:2019} revealed that spatially heterogeneous distal ejecta deposition plays an important role in explaining the observed lunar crater distribution~\citep{Minton:2019}. Statistical analysis of crater distribution is certainly useful for discussing the surface processes occurring on planetary surfaces. In addition to these statistical approaches, microscopic physical processes governing the individual relaxation of craters should also be studied. By combining both approaches, we can develop a proper understanding of general crater degradation processes. In this study, we concentrate on the latter. In particular, we focus on the crater-wall relaxation due to the accumulation of numerous small-scale impacts. A schematic diagram of a small-scale impact on a large-scale crater is shown in Fig.~\ref{fig:setup}(a). Through repeated small-scale impacts, a large-scale crater shape can be gradually degraded. 

Gradual crater-wall relaxation due to small-scale impacts is a crucial process on relatively large target bodies such as the Moon or Mars. Indeed, this process has been applied to the analysis of lunar craters~\citep{Fassett:2014}. For small bodies such as asteroids, in contrast, global seismic shaking is easily triggered and efficiently degrades craters. The fundamental process of crater degradation by global seismic shaking can be modeled using a nonlinear transport law~\citep{Roering:1999, Roering:2001,Tsuji:2018,Tsuji:2019}. However, details on crater degradation due to small-scale impacts are not yet understood well. Impact-induced slope relaxation requires investigation as a fundamental process of crater degradation through small-scale impacts. However, impact-induced landsliding has only been experimentally studied under microgravity conditions to mimic small-body conditions~\citep{Hofmann:2017}. Relaxation of sloped terrain by impacts should be investigated in a wider variety of conditions because this process is more relevant to large target bodies than microgravity conditions. 

Furthermore, only the asymmetric-ejecta-deposition effect has been theoretically considered in the model of slope relaxation by small-scale impacts \citep{Soderblom:1970}. To quantitatively model crater-wall relaxation through the accumulation of small-scale impacts, the relaxation of the local slope due to the downward migration of the centroid of the impacted slope should be appropriately understood. Therefore, the downhill migration of ejecta caused by small-scale impacts was considered by \citet{Soderblom:1970}. Hence, the migration of the centroid driven by small-scale impacts is considered in characterizing the slope relaxation in this study. \citet{Soderblom:1970} obtained a relation between the target slope angle $\theta$, diameter of the crater cavity produced by a small-scale impact $D_c$, and resultant centroid migration per impact in the downhill direction $x_\mathrm{mig}$, as 
\begin{equation}
x_\mathrm{mig} = \frac{D_c}{4}\frac{(\beta+1)(\beta-1)}{\beta+2} \cot \alpha \tan \theta ,
\label{eq:Soderblom_original}
\end{equation}
where $\alpha$ and $\beta$ are the apex angle of the ejecta cone and a parameter for the ejecta deposition profile, respectively. With reasonable assumptions $\beta \simeq 3$--$4$ and $\alpha=45^{\circ}$, \citet{Soderblom:1970} derived an approximated form,
\begin{equation}
x_\mathrm{mig} \simeq \frac{D_c}{2} \tan\theta. 
\label{eq:Soderblom}
\end{equation}
This downward centroid migration is caused by asymmetric ejecta deposition due to the presence of gravity, resulting in effective slope relaxation. While equation~(\ref{eq:Soderblom}) is reasonably simple, the validity of this relation has not been experimentally examined. Therefore, we analyze the experimental data of \citet{Takizawa:2020} in which a solid projectile was impacted onto an inclined granular surface at various incident angles. From the measured crater morphologies, we evaluate the parameter dependence of the migration of the centroid caused by the impact on an inclined surface. Subsequently, the validity of equation~(\ref{eq:Soderblom}) can be assessed.   

As most solar-system bodies are covered by regolith, granular mechanics should be considered to properly model the surface terrain behaviors. For example, \citet{Murdoch:2013,Maurel:2016,Cheng:2019} discussed various granular mechanics in the context of planetary surface processes. In this study, we also employ granular matter (dry sand) as the target material to consider the planetary slope relaxation.

\section{Experiment}
\begin{figure}
	\includegraphics[width=\columnwidth]{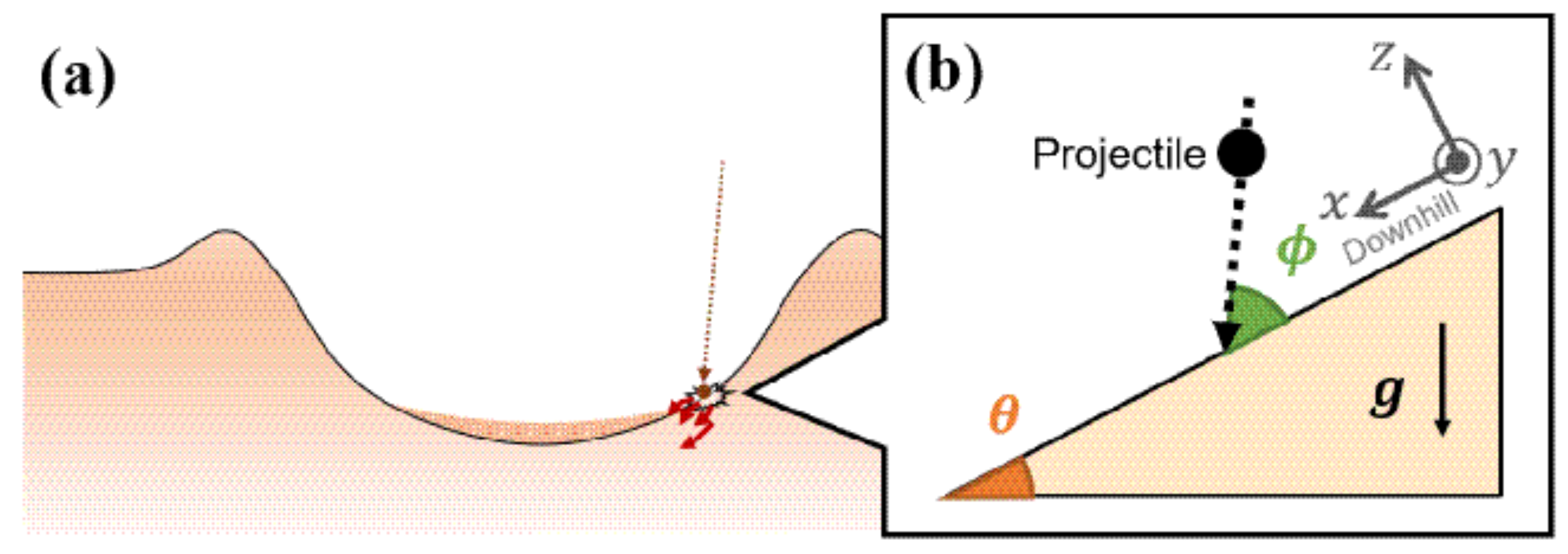}
  \caption{Schematic images of (a)~the considered scenario and (b)~the experimental setup. A tiny impact occurring on a large-scale crater and its associated crater-wall relaxation are considered. Definitions of the coordinate system, inclination angle $\theta$, and incident angle $\phi$ are presented. The variation ranges of the angles are $0\leq \theta \leq 33^{\circ}$ and $50 \leq \phi \leq 130^{\circ}$. Details of the experimental setup can be found in \citet{Takizawa:2020in}. }
    \label{fig:setup}
\end{figure}

In this study, experimental data taken by an instrument recently developed for oblique impact experiments are analyzed. Here, we summarize the concept of the instrument. Details of the experimental system can be found in \citet{Takizawa:2020in}. A schematic diagram of the considered scenario is shown in Fig.~\ref{fig:setup}(b). In the experimental setup, both the inclination angle of the target layer $\theta$ and incident angle $\phi$ can be varied. The common rotational axis is set as the $y$-direction. A high-speed camera films the impact to identify the impact point. The direction perpendicular to the target surface is taken as the $z$-direction. The vertically upward direction is defined as positive, and $z=0$ corresponds to the target surface level before impact. The positive direction of the $x$-axis corresponds to the downhill direction of the slope. Toyoura sand (of grain diameter range 0.1--0.3\,mm and angle of repose 34$^{\circ}$) is used as the target sand material. The packing fraction and material true density of the sand are $0.55$ and $2.5 \times 10^3$\,kg\,m$^{-3}$, respectively. A plastic projectile of diameter $6$\,mm with a mass of $0.12$, $0.25$, or $0.4$\,g is impinged onto the target sand layer with an impact speed of $7 \leq v_\mathrm{imp} \leq 97$\,m\,s$^{-1}$. 

The resultant crater shapes are measured using a laser profilometry system. The resolutions of this measurement in the $x$-, $y$-, and $z$-directions are $dx=dy=50$\,$\mu$m, and $dz=0.5$\,$\mu$m, respectively. The field of measurement spans $191{\times}65$\,mm$^{2}$ in the $x$$y$ plane. From the measured three-dimensional (3D) crater-shape data, we compute the centroid migration. The data used in this study were also analyzed by \citet{Takizawa:2020} and used to derive scaling laws for the crater dimensions. Using the same data, we focus here on the asymmetry of craters formed on an inclined surface. It should be noted that the considered scales are distinct between these two analyses. In the former~\citep{Takizawa:2020}, experimentally produced craters were considered the contraction models of actual large-scale craters. In the latter (this study), in contrast, we regard the inclined target layer as a part of the large-scale crater's wall, as shown in Fig.~\ref{fig:setup}(a). Some data used in \citet{Takizawa:2020} do not have sufficient quality for discussion of the slope relaxation. Thus, we use the data from 56 of the 98 impacts reported in \citet{Takizawa:2020}. 

\section{Results and analyses}
\subsection{Crater shapes}
\begin{figure}
	\includegraphics[width=\columnwidth]{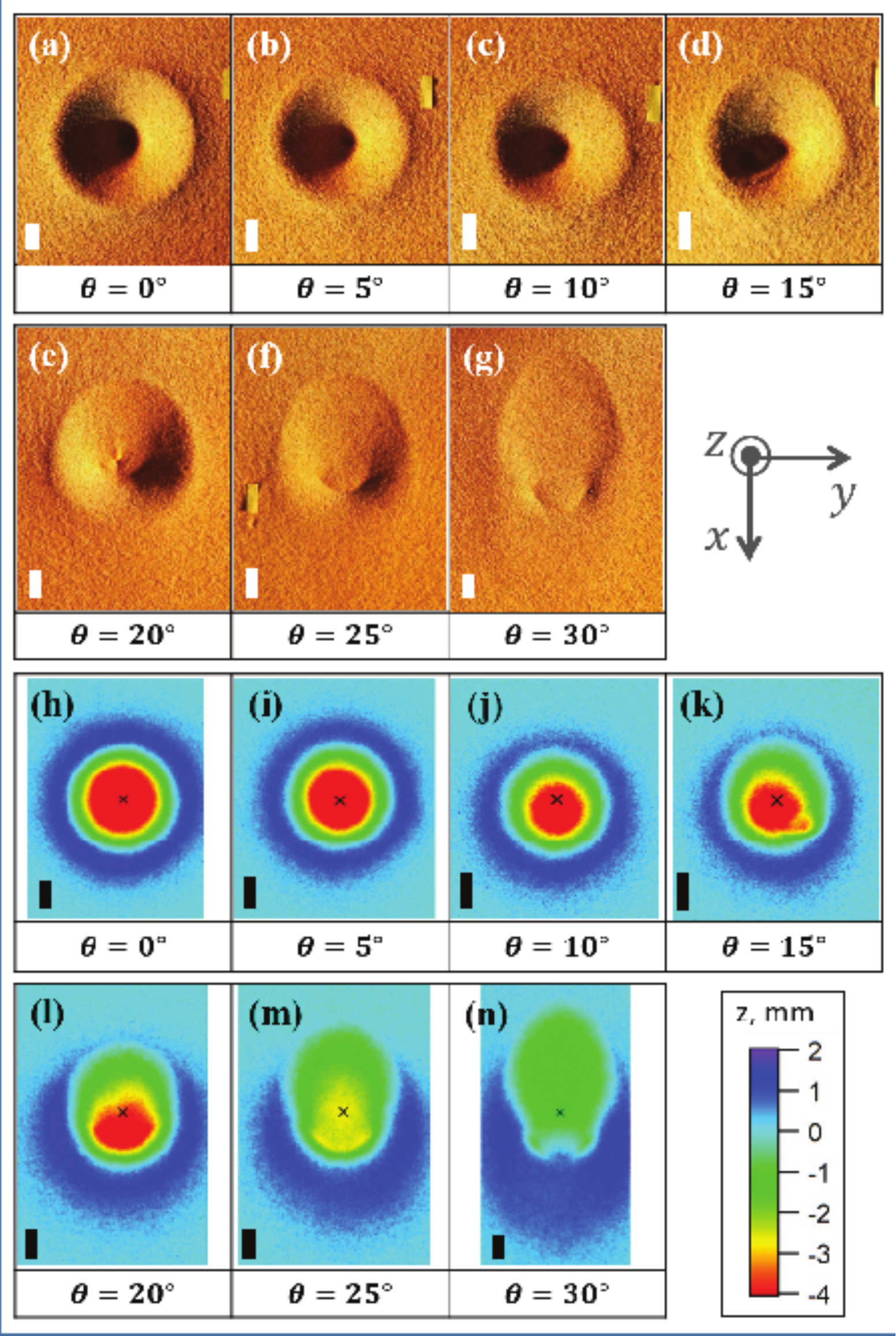}
    \caption{Example data of craters produced on inclined surfaces. All craters shown here are produced by perpendicular impacts ($\phi=90^{\circ}$). (a--g) Top-view photos of the resultant crater shapes. (h--n) Corresponding height maps measured by the laser profilometry system. The spatial resolution of the profile data is $50$~$\mu$m in the $x$- and $y$-directions and $0.5$~$\mu$m in the $z$-direction. Scale bars indicate $10$~mm. A symmetric crater is formed for $\theta=0^{\circ}$, and the asymmetry of the crater shapes gradually increases as $\theta$ increases. In particular, the transient crater cavity is almost buried by the landsliding of crater-wall collapse for $\theta=30^{\circ}$. The black crosses shown in (h--n) indicate the impact points. }
    \label{fig:raw_prof}
\end{figure}

Examples of the raw crater shape data are shown in Fig.~\ref{fig:raw_prof}. Top-view images (Fig.~\ref{fig:raw_prof}(a--g)) and 3D height maps $z(x_{i},y_{j})$ (Fig.~\ref{fig:raw_prof}(h--n)) of various target inclinations $\theta$ are presented, where $i$ and $j$ are position indexes; $x_{i+1}-x_{i}=dx$ and $y_{j+1}-y_{j}=dy$. In these impacts, the projectile impacts perpendicularly onto a target surface with $\phi=90^{\circ}$. A symmetric crater shape can be observed in Fig.~\ref{fig:raw_prof}(a,h) because of the complete symmetry in the vertical impact onto a horizontal target. A similar trend can be confirmed for $\theta=5^{\circ}$ (Fig.~\ref{fig:raw_prof}(b,i)). In the range $10^{\circ} \leq \theta \leq 15^{\circ}$, although the crater shapes appear roughly symmetric (Fig.~\ref{fig:raw_prof}(c,d)), asymmetry of ejecta deposition can be observed, as shown in Fig.~\ref{fig:raw_prof}(j,k). When $\theta$ is greater than or equal to $20^{\circ}$, asymmetric crater shapes can be clearly observed. In this regime, both the ejecta deposition and the collapse of the crater wall contribute to asymmetric crater modification (Fig.~\ref{fig:raw_prof}(e--g, l--n)). Because of these processes, the centroid of the resultant crater migrates. Usually, the centroid migrates downhill because gravity is a principal source of asymmetry. In other words, the slope is effectively relaxed by gravity in various ways. To quantify the slope relaxation, dependences of the centroid migration on the impact conditions must be measured. 

\subsection{Centroid migration measurement method}
\begin{figure*}
	\includegraphics[width=2.0\columnwidth]{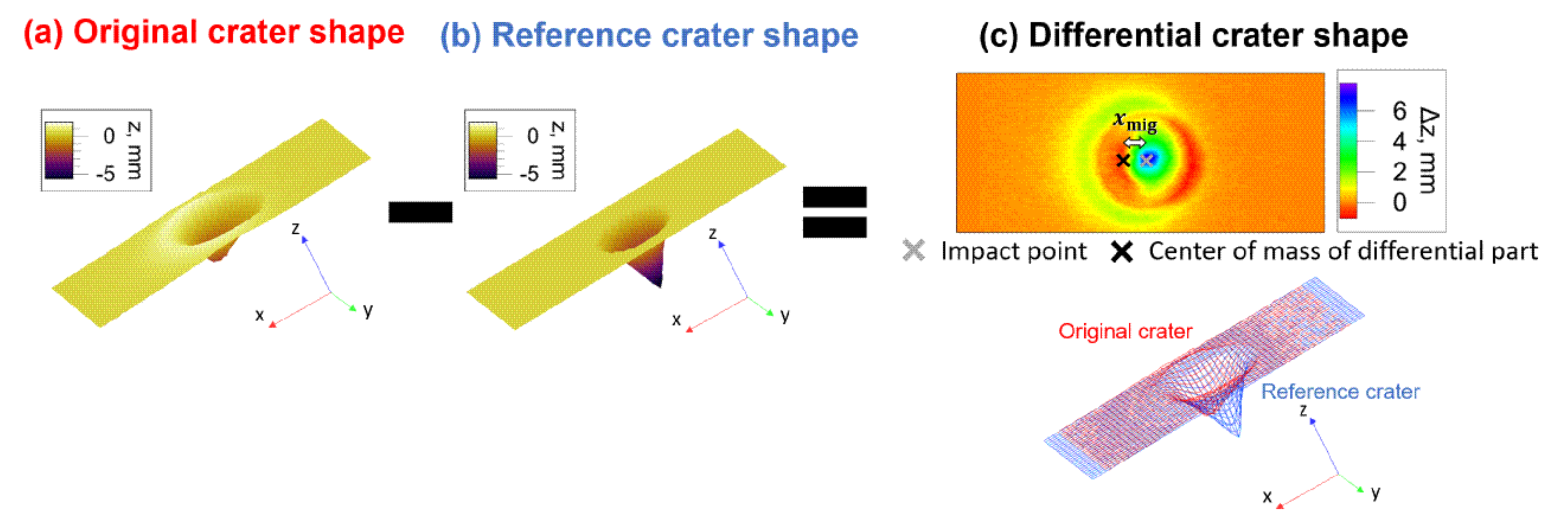}
  \caption{Computation example of centroid migration. 3D maps of (a) original crater shape and (b) (scaled) reference crater shape. (c) Height map of the differential crater shape (upper) and direct comparison of the original and (scaled) reference crater shapes (lower). The reference crater shape is defined by the crater-cavity profile (without rim) produced by the impact of $\theta=0^{\circ}$ and $\phi=90^{\circ}$. The centroid migration distance $x_\mathrm{mig}$ is computed using equation~(\ref{eq:centroid}). In the 3D plots, the depth values are exaggerated.} 
    \label{fig:method}
\end{figure*}

Using the 3D height maps, we compute the centroid migration distance $x_\mathrm{mig}$, as shown in Fig.~\ref{fig:method}. First, the cavity shape of $\theta=0^{\circ}$ and $\phi=90^{\circ}$ is defined as a reference shape because this profile represents the ideally symmetric situation. It should be noted that, to define the reference shape, the crater rim is removed from the original crater shape. Because we aim to characterize the asymmetry of ejecta deposition and crater-wall collapse, the distribution of ejected/drifted materials should be measured. Therefore, the symmetrically excavated cavity shape should be taken as the baseline (bottom) for calculating the mass redistribution. Next, we have to expand (scale up) or reduce (scale down) the reference shape for comparison with various crater shapes because the crater-cavity size depends on the impact conditions. For this purpose, we modify the scale of the reference shape using the scaling laws  reported in \citet{Takizawa:2020}. Basically, the reference shape is rescaled so that the crater width in the $y$-direction $D_\mathrm{cy}$ of the reference crater shape becomes identical to that of the original crater shape. The position of the reference shape is then adjusted to standardize the impact point. Subsequently, the differential height map $z_\mathrm{dif}(x_{i},y_{j})$ is computed by subtracting the scaled reference shape $z_\mathrm{ref}(x_{i},y_{j})$ from the original profile $z(x_{i},y_{j})$; i.e., $z_\mathrm{dif}(x_{i},y_{j})=z(x_{i},y_{j})-z_\mathrm{ref}(x_{i},y_{j})$. An example of this procedure for $\theta=20^{\circ}$ and $\phi=90^{\circ}$ is shown in Fig.~\ref{fig:method}. Assuming uniform density of the target sand layer, the centroid migration distance $x_\mathrm{mig}$ is computed from $z_\mathrm{dif}(x_{i},y_{j})$ and $z_\mathrm{ref}(x_{i},y_{j})$ as
\begin{equation}
x_\mathrm{mig} = \frac{\sum_{j=y_\mathrm{min}}^{y_\mathrm{max}} \sum_{i=x_\mathrm{min}}^{x_\mathrm{max}} \left(x_{i} - x_\mathrm{imp}\right) z_\mathrm{dif}(x_{i},y_{j})  }{ \sum_{j=y_\mathrm{min}}^{y_\mathrm{max}} \sum_{i=x_\mathrm{min}}^{x_\mathrm{max}} \left| z_\mathrm{ref}(x_{i},y_{j}) \right| }. 
\label{eq:centroid}
\end{equation}
where $x_\mathrm{imp}$ denotes the $x$-component of the impact point, which is determined by inspection of the video data taken by the high-speed camera~\citep{Takizawa:2020in}. We do not compute the migration in the $y$-direction $y_\mathrm{mig}$ because it must be negligibly small owing to the symmetry. Moreover, $y_\mathrm{mig}$ does not contribute to the relaxation process. Thus, $y_\mathrm{mig}$ is not required in this study. 
$x_\textrm{min}$, $x_\mathrm{max}$, $y_\mathrm{min}$, and $y_\mathrm{max}$ are determined to properly measure the crater shape and the corresponding $x_\mathrm{mig}$. 
Owing to the asymmetric ejecta deposition and crater-wall collapse, clear migration of the centroid $x_\mathrm{mig}$ is confirmed in Fig.~\ref{fig:method}. This procedure is applied to all crater profiles to derive their $x_\mathrm{mig}$ values.

\subsection{Centroid migration for inclined target and oblique impact}
\begin{figure*}
	\includegraphics[width=1.6\columnwidth]{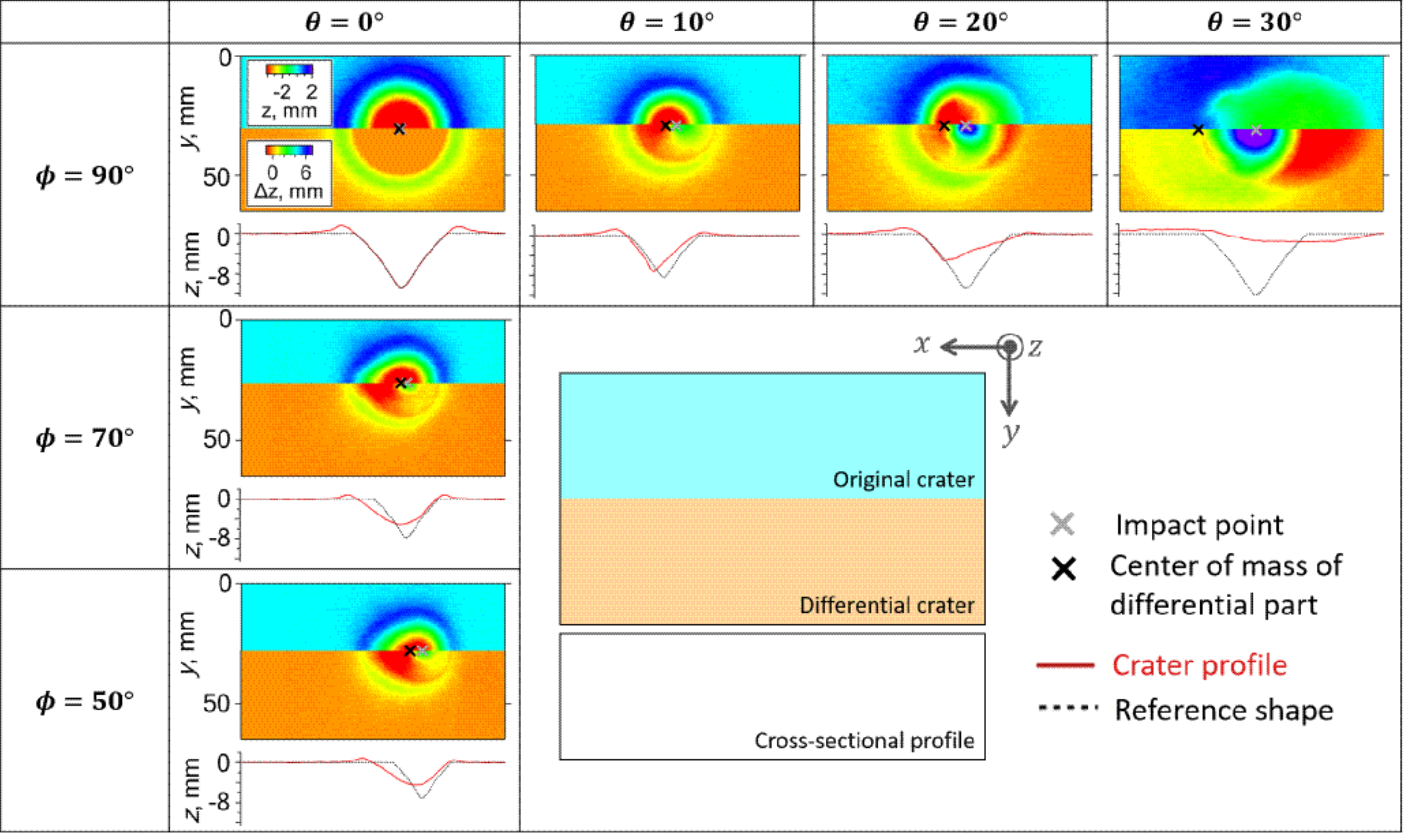}
    \caption{(Original and differential) crater height maps and (original and reference) cross-sectional profiles of various inclination angles $\theta$ and incident angles $\phi$. The upper and lower halves of the height maps correspond to the original and differential shapes, respectively. The gray and black crosses indicate the impact point and the migrated centroid position, respectively. The red curves in the cross-sectional profiles denote the original crater shapes at the central position. The black dotted curves are the corresponding (scaled) reference shapes.}
    \label{fig:vertical_results}
\end{figure*}

Typical analyzed results for various inclination angles are presented in Fig.~\ref{fig:vertical_results}. The upper and lower halves of the height maps in Fig.~\ref{fig:vertical_results} show the original profiles $z(x_{i},y_{j})$ and differential profiles $z_\mathrm{dif}(x_{i},y_{j})$, respectively. The bottom row displays the cross-sectional profiles $z(x_{i},y_\mathrm{imp})$ and corresponding reference profiles $z_\mathrm{ref}(x_{i},y_\mathrm{imp})$, where $y_\mathrm{imp}$ is the $y$-component of the impact point. The gray and black crosses in Fig.~\ref{fig:vertical_results} indicate the impact point $(x_\mathrm{imp},y_\mathrm{imp})$ and the centroid position computed by equation~(\ref{eq:centroid}) $(x_\mathrm{imp}+x_\mathrm{mig},y_\mathrm{imp})$, respectively. It can be confirmed that $x_\mathrm{mig}$ increases as $\theta$ increases. For large-$\theta$ cases, a large fraction of the crater cavity is buried by landsliding of the crater-wall collapse. 

The effect of oblique impact on centroid migration is also shown in Fig.~\ref{fig:vertical_results}. Because of the obliquity of impact, the crater shape becomes asymmetric and egg-like when $\phi=50^{\circ}$.  This asymmetric excavation results in the migration of the centroid. However, the effect of oblique impact is not significant compared with that of the target inclination. This means that the crater-wall collapse is the most crucial process for evaluating centroid migration. The effects of asymmetric ejecta deposition and asymmetric excavation have relatively minor impacts. 

\begin{figure}
	\includegraphics[width=\columnwidth]{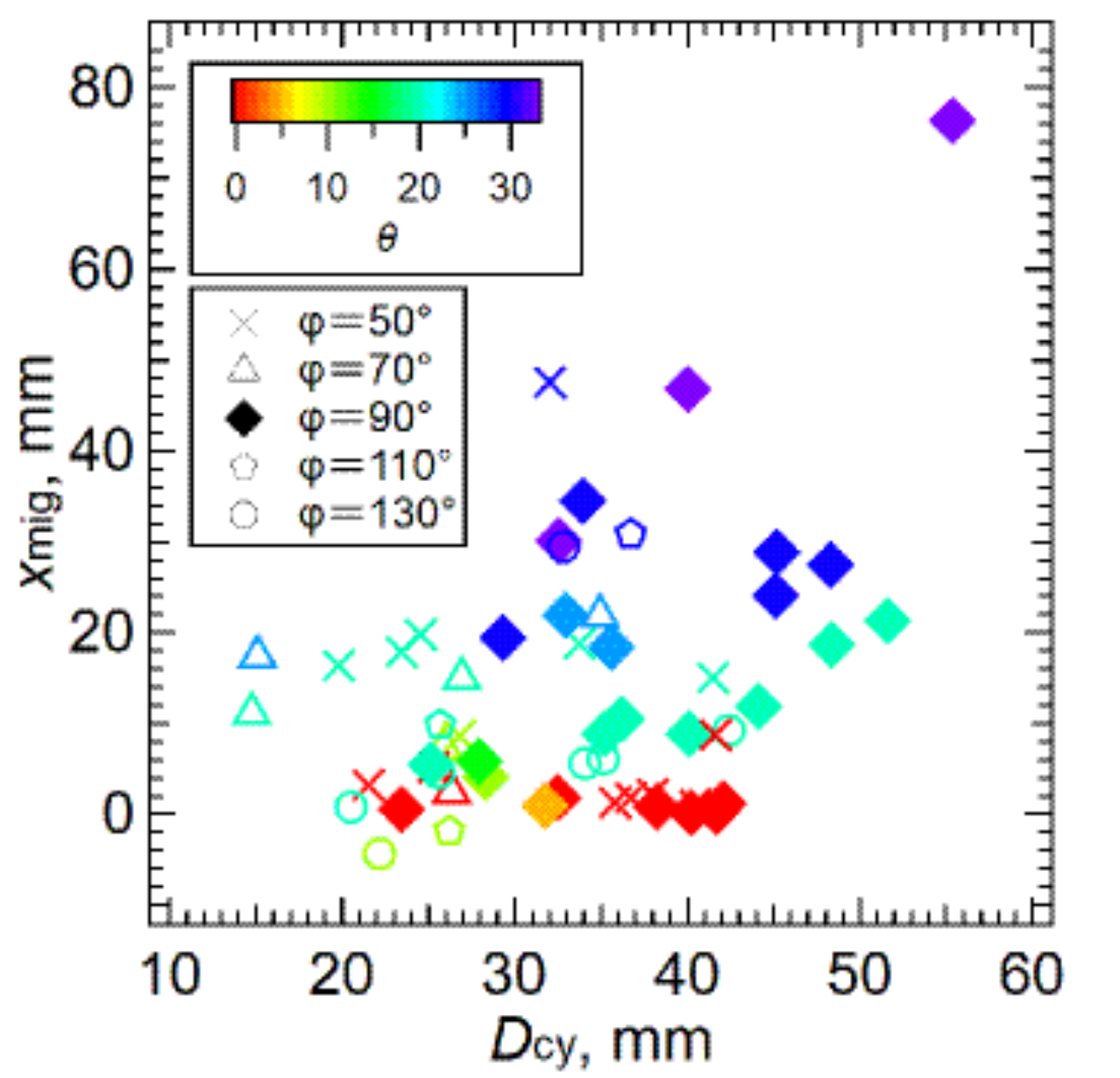}
  \caption{Parameter dependences of the measured centroid migration $x_\mathrm{mig}$ computed by equation~\eqref{eq:centroid}. The inclination angle of the target surface and oblique-impact angle are represented by color and symbol variations, respectively. }
    \label{fig:x_mig_raw}
\end{figure}

In Fig.~\ref{fig:x_mig_raw}, all the measured $x_\mathrm{mig}$ data are plotted as a function of $D_\mathrm{cy}$. The colors and symbols in Fig.~\ref{fig:x_mig_raw} indicate $\theta$ and $\phi$, respectively. As can be seen, $x_\mathrm{mig}$ increases as $\theta$ increases. When the target inclination is small ($\theta \simeq 0$), the centroid migration is also small ($x_\mathrm{mig} \simeq 0$) regardless of the $D_\mathrm{cy}$ and $\phi$ values. In the large-$\theta$ range, a positive correlation between $x_\mathrm{mig}$ and $D_\mathrm{cy}$ can be confirmed. This size dependence suggests that the normalized migration distance $x_\mathrm{mig}/D_\mathrm{cy}$ should be analyzed. Regarding the effect of $\phi$, a slight negative correlation between $x_\mathrm{mig}$ and $\phi$ can be observed, particularly in the $\theta \simeq 20^{\circ}$ regime. However, the number of oblique impacts is limited due to the technical difficulty of shallow-impact experiments (e.g., ricochet, impact point identification, etc.). Thus, in the following, we first analyze the data of $\phi=90^{\circ}$ to develop a physical model. The model is then applied to various $\phi$ cases. Specifically, in the next section, we introduce a nonlinear model for centroid migration that includes three effects: crater-wall collapse, asymmetric ejecta deposition, and asymmetric excavation. 

\subsection{Nonlinear migration model}
\begin{figure}
	\includegraphics[width=\columnwidth]{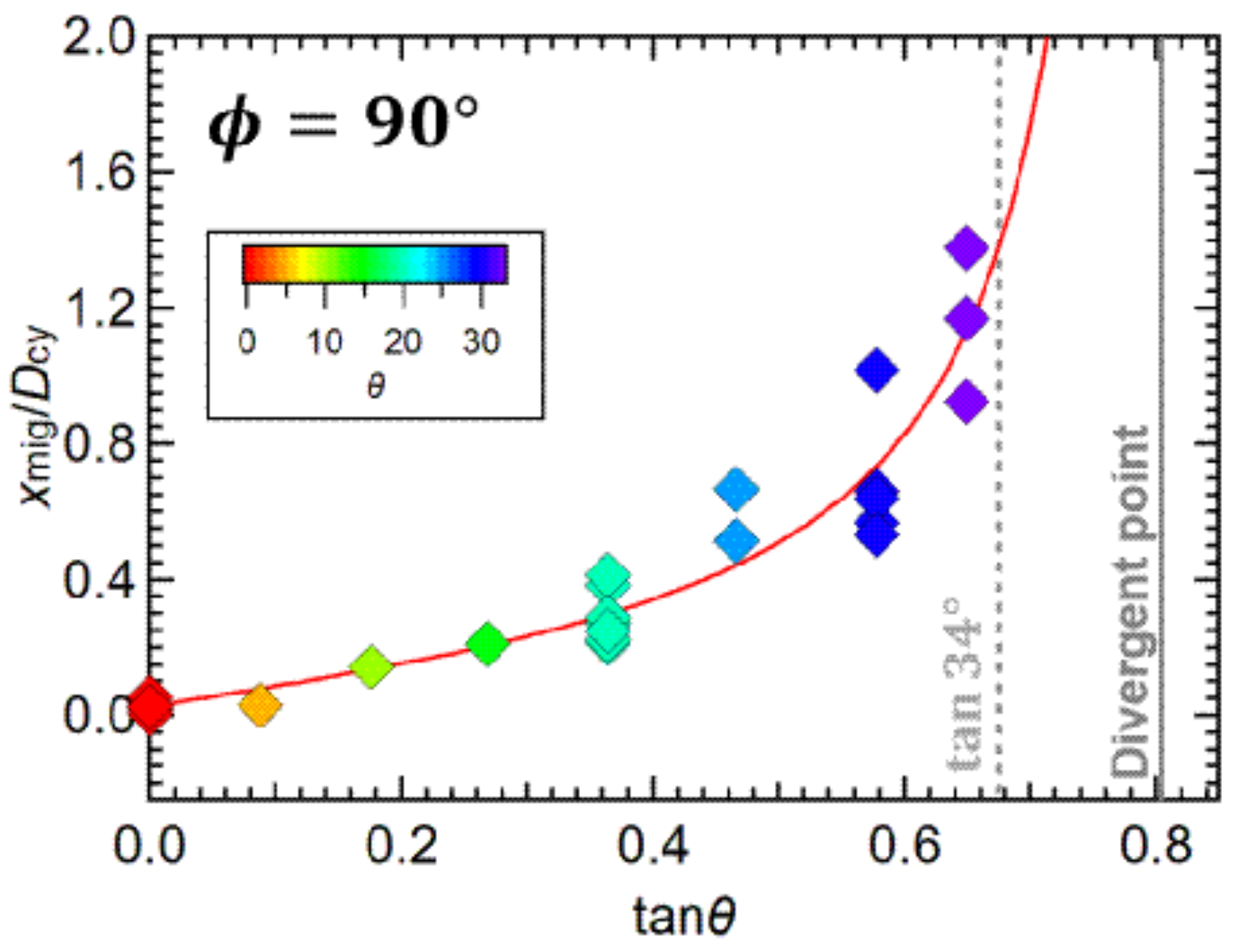}
  \caption{Normalized centroid migration $x_\mathrm{mig}/D_\mathrm{cy}$ vs.~$\tan\theta$ for perpendicular impacts ($\phi=90^{\circ}$). The solid curve indicates the fitting to equation~(\ref{eq:nonlinear_law}) with $K=0.6$, $\mu=0.80$, and $c(\phi=90^{\circ})=0.03$. The vertical lines denote $\tan 34^{\circ}$ (angle of repose) and $\tan\theta=\mu$ (divergent point). The color code is identical to that used in Fig.~\ref{fig:x_mig_raw}. }
    \label{fig:relax_vertical}
\end{figure}

To quantitatively analyze centroid migration, $x_\mathrm{mig}$ is normalized to the crater's minor-axis ($y$-axis) diameter $D_\mathrm{cy}$. The relation between $x_\mathrm{mig}/D_\mathrm{cy}$ and the initial inclination $\tan\theta$ for perpendicular impacts ($\phi=90^{\circ}$) is plotted in Fig.~\ref{fig:relax_vertical}, and is clearly nonlinear. This implies that the conventional form proposed by \citet{Soderblom:1970} (equation~(\ref{eq:Soderblom})) cannot be simply applied. To improve the model, we borrow an idea from the nonlinear transport law proposed for a vibrated hillslope relaxation~\citep{Roering:1999,Roering:2001}. Specifically, we introduce a model, 
\begin{equation}
\frac{x_\mathrm{mig}}{D_\mathrm{cy}} = K \left( \frac{\tan \theta}{1-(\tan\theta/\mu)^2} \right) + c(\phi),
\label{eq:nonlinear_law}
\end{equation}
where $K$, $\mu$, and $c(\phi)$ are fitting parameters. While this type of nonlinear model was initially considered to explain landsliding dynamics~\citep{Roering:1999,Roering:2001} rather than ejecta deposition, the consistency between the model of equation~\eqref{eq:nonlinear_law} and the ejecta deposition model~\citep{Soderblom:1970} will be discussed later.

First, we analyze the perpendicular impacts. The solid curve shown in Fig.~\ref{fig:relax_vertical} indicates the least-squares fitting by equation~(\ref{eq:nonlinear_law}). The fitting curve successfully captures the data behavior. The estimated fitting parameter values are $K=0.6 \pm 0.1$, $0.80 \pm 0.04$, and $c(\phi=90^{\circ})=0.03 \pm 0.04$. Note that the value of $c(\phi=90^{\circ})$ is approximately zero. Thus, only the first term on the right-hand side of equation~\eqref{eq:nonlinear_law} is necessary to model the perpendicular-impact data.  

\begin{figure}
	\includegraphics[width=\columnwidth]{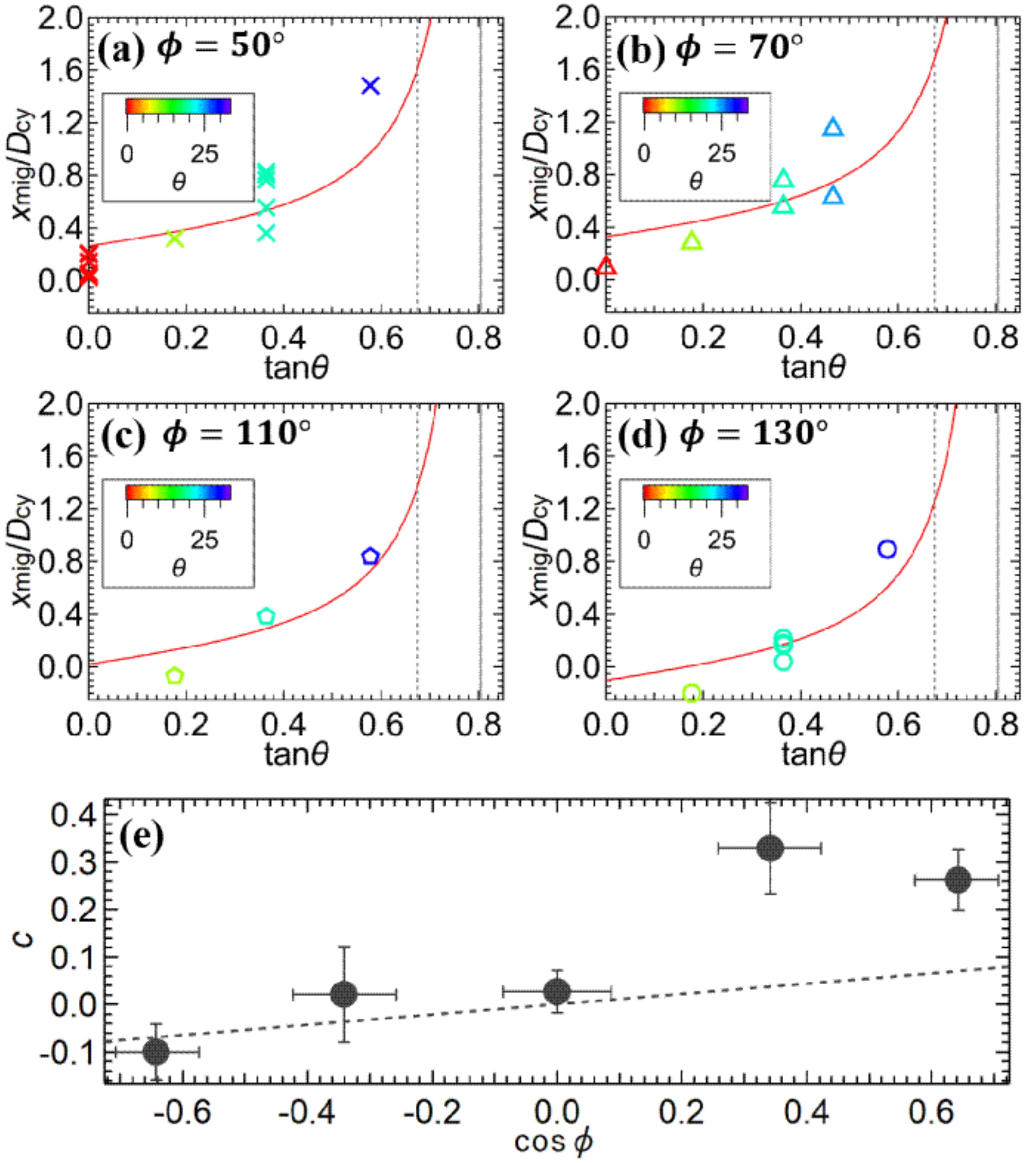}
  \caption{Normalized centroid migration $x_\mathrm{mig}/D_\mathrm{cy}$ vs.~$\tan\theta$ for oblique impacts.  (a--d) Measured $x_\mathrm{mig}/D_\mathrm{cy}$ results with corresponding fitting curves. The parameter values of $K(=0.6)$ and $\mu(=0.80)$ are fixed, and only $c(\phi)$ is a free fitting parameter. (d) Measured $c(\phi)$ values plotted as a function of $\cos\phi$. The dashed line indicates the empirical fitting, $c(\phi)=c_\mathrm{0}\cos\phi$ with $c_\mathrm{0}=0.2$. The color code is identical to that used in Fig.~\ref{fig:x_mig_raw}. }
    \label{fig:relax_oblique}
\end{figure}

Next, the effect of oblique impact is analyzed to evaluate the role of the term $c(\phi)$. The number of data points for oblique impact is actually limited. Thus, we fix the values of $K~(=0.6)$ and $\mu~(=0.80)$ in the fitting; i.e., only $c(\phi)$ is a free fitting parameter. The reason for this fitting procedure is explained later based on the physical meaning of the model equation~(\ref{eq:nonlinear_law}). The $x_\mathrm{mig}/D_\mathrm{cy}$ data and the corresponding fitting results for various $\phi$ cases are shown in Fig.~\ref{fig:relax_oblique}(a--d). Although the number of data is limited and considerable data scattering can be observed, the fitting curves adequately reproduce the global trend of the data. Due to the limited data, the nonlinearity in Fig.~\ref{fig:relax_oblique}(a--d) is not very conclusive. In particular, possible linear trends can be observed in Fig.~\ref{fig:relax_oblique}(c, d). Negative $x_\mathrm{mig}/D_\mathrm{cy}$ in the small-$\theta$ regime implies upward migration of the centroid. As the simplest way to account for this effect, we introduce the parameter $c(\phi)$. That is, we assume that the nonlinear $\tan\theta$ dependence of $x_\mathrm{mig}/D_\mathrm{cy}$ obtained for the perpendicular impact is also applicable to oblique impacts.  The obtained values of $c(\phi)$ are plotted as a function of $\cos \phi$ in Fig.~\ref{fig:relax_oblique}(e). While the range of $c(\phi)$ variation is limited, a systematic behavior of $c(\phi)$ can be confirmed. Here, we empirically fit $c(\phi)$ to a simple relation, $c(\phi) = c_\mathrm{0} \cos \phi$  (in the $\cos \phi \leq 0$ regime). The value of the obtained fitting coefficient is $c_\mathrm{0}=0.2$. In the positive $\cos\phi$ regime, the $c(\phi)$ values exceed the expected values. This asymmetric behavior of $c(\phi)$ originates from the relation between oblique impact and target inclination. When both effects increase the centroid migration ($\theta >0$ and $\cos\phi >0$), $c(\phi)$ becomes large. When they work oppositely ($\theta >0$ and $\cos\phi<0$), $c(\phi)$ becomes relatively small. Therefore, $c(\cos\phi)$ appears nonlinear in Fig.~\ref{fig:relax_oblique}(e).   

This modeling is approximately empirical. However, the form of nonlinearity in equation~\eqref{eq:nonlinear_law} can be physically understood by considering the frictional dissipation and grain agitation due to the energy injection~\citep{Tsuji:2018,Tsuji:2019}. Therefore, in this study, we employ and improve a physically understandable model to avoid unnecessarily complex modeling.

\section{Discussion}
The form of the nonlinear centroid migration model (equation~(\ref{eq:nonlinear_law})) must be related to the physical processes of the crater-shape modification. In fact, a similar model can explain the relaxation of a vibrated granular heap~\citep{Roering:1999,Roering:2001,Tsuji:2018,Tsuji:2019}. In particular, \citet{Tsuji:2018,Tsuji:2019} revealed the physical mechanisms and parameter dependences of the model in detail. Therefore, we follow the idea of this model. The most important parameter in equation~(\ref{eq:nonlinear_law}) is $\mu$, which corresponds to the diverging point of $x_\mathrm{mig}/D_\mathrm{cy}$ (Figs.~\ref{fig:relax_vertical}, \ref{fig:relax_oblique}). Namely, $\tan\theta$ cannot exceed $\mu$. This upper limit of $\tan\theta$ can be regarded as an effective friction coefficient. Indeed, the corresponding friction angle $\arctan \mu = 40^{\circ}$ is slightly larger than but close to the angle of repose of the target sand $34^{\circ}$. Thus, we consider this value to represent the effective friction coefficient characterizing the landsliding of the crater-wall collapse. Because the nonlinear effect is enhanced as $\tan\theta$ approaches $\mu$, a significant increase in $x_\mathrm{mig}/D_\mathrm{cy}$ is observed in the large-$\tan\theta$ regime. In other words, the instability of the steep granular heap is the origin of the nonlinearity. For small $\theta$, equation~(\ref{eq:nonlinear_law}) can be linearized as $x_\mathrm{mig}/D_\mathrm{cy} = K \tan\theta+c(\phi)$. This form is consistent with equation~(\ref{eq:Soderblom}) because the obtained values  are $K=0.6 \simeq 1/2$ and $c(\phi=90^{\circ})=0.03 \pm 0.04 \simeq 0$. When $\theta$ is small, asymmetric ejecta deposition dominates the centroid migration, as modeled by \citet{Soderblom:1970}. Therefore, it can be said that the parameter $K$ characterizes the asymmetric ejecta deposition effect due to the inclination of the surface relative to the gravitational direction. Finally, the effect of the oblique impact $c(\phi)$ is independently added to the form. Because the variation range of $c(\phi)$ is limited, this effect is not very significant. However, the lateral momentum transfer driven by oblique impact certainly affects centroid migration. This effect should be independent of other effects. Thus, we fix $K$ and $\mu$ values in the fittings shown in Fig.~\ref{fig:relax_oblique}(a--d). The nonlinearity seen in $c(\cos\phi)$ originates from the asymmetric relation between $\theta$ and $\phi$; $\phi<90^{\circ}$ and $\phi>90^{\circ}$  indicate the impact from the upward and downward directions of the slope, respectively. In summary, model parameters $\mu$, $K$, and $c(\phi)$ represent the effective friction coefficients governing landsliding, the effect of asymmetric ejecta deposition, and the effect of oblique impact, respectively. 

Although the experimental data are consistent with the model, the number of data is not sufficient, particularly for oblique impacts. However, to consider the actual crater degradation, oblique impacts are important. By considering the impact angle population~\citep{Ross:1968}, oblique impacts with $\phi>90^{\circ}$ could be frequent on inclined crater walls. Therefore, we should exercise caution in the modeling of the impact obliquity effect. In fact, the effect of impact obliquity is not very significant, as discussed above. This tendency is consistent with a previous study on crater morphology by oblique impacts~\citep{Gault:1978}. Overall, we believe that the proposed model provides a reasonable approximation. To obtain a precise model, however, more detailed experiments with wider parameter ranges are necessary.

The form of equation~(\ref{eq:nonlinear_law}) is free from the gravity effect. The migration distance of the centroid is primarily determined by the geometric conditions of the target terrain and impact. The values of $\mu$, $K$, and $c(\phi)$ can be regarded as constants when the target surface consists of a simple regolith (dry sand) layer. The scale of impact inertia (momentum/energy) and gravity determine the size of the crater $D_\mathrm{cy}$. That is, the effect of gravity is included in the estimate of $D_\mathrm{cy}$. Such scaling laws for $D_\mathrm{cy}$ and other crater dimensions have already been obtained by \citet{Takizawa:2020}. Thus, the actual centroid migration distance per impact event can be estimated by combining equation~(\ref{eq:nonlinear_law}) and the result of \citet{Takizawa:2020}. 

According to equation~(\ref{eq:nonlinear_law}), steep slopes are relaxed significantly faster than gentle slopes. This feature is similar to in slope relaxation driven by seismic shaking because the functional form of slope relaxation is identical between these two processes. The challenge is to determine which one (seismic shaking or crater-wall collapse) is dominant in the actual degradation process of craters. The answer to this question depends on the size of the target body and the collision-frequency distribution of the impactors.

To evaluate the age of craters degraded by the accumulation of impact-induced slope relaxation, information on the size-dependent collision-frequency distribution on the surface of the target body is required. In addition to the classical method used in \citet{Soderblom:1970}, small-body distribution and their collision-frequency models were developed recently~\citep[e.g.,][]{Bottke:1993,Bottke:1994a,Bottke:1994b,OBrien:2005,Richardson:2005,Yamada:2016}. By combining these collision models and the current results, a new chronology for crater degradation can be established. 

As a first-order approximation, the classical model proposed by \citet{Soderblom:1970} provides a reasonable estimate because the current experimental result supports the model, particularly in the small-$\tan\theta$ regime. This paper reports experimental evidence of the validity of equation~(\ref{eq:Soderblom}) for the first time. As mentioned before, the steep slope is rapidly relaxed owing to the nonlinear nature of equation~(\ref{eq:nonlinear_law}). Thus, the actual crater relaxation timescale is primarily determined by the slow (linear) relaxation of gentle slopes. The linear approximation results in the linear diffusional relaxation of the crater shape~\citep{Soderblom:1970}. This linear approximation enables us to analytically compute various quantities. However, the nonlinear law (equation~(\ref{eq:nonlinear_law})) makes it impossible. The linear approximation is sufficient for most cases, whereas the nonlinear effect must be considered to analyze the degradation of relatively young surfaces. 

Finally, the limitation of the impact experiment performed by \citet{Takizawa:2020} that provides the data analyzed in this study should be discussed. In the experiment, the impact velocity was less than 100~m~s$^{-1}$, which was slower than the sound velocity of the target. In addition, the experiments were performed under atmospheric conditions. Thus, the fundamental physical process of the impact could be different from those of high-speed impacts occurring on the surfaces of airless objects such as the Moon. However, the scaling laws obtained in \citet{Takizawa:2020} are consistent with the results of the high-speed impact experiment performed by~\citet{Schmidt:1980}. Therefore, we consider that the experimental result can also be applied to planetary scenarios. A more serious problem could be the scale independence of the current model. Our model (equation~\eqref{eq:nonlinear_law}) is written in dimensionless form. Based on the concept of dimensional analysis, such a relation can be applied to large-scale phenomena for which the governing physical mechanism is identical. However, a recent study on lunar crater degradation~\citep{Basilevsky:2018,Minton:2019} highlighted the possible scale dependence of crater degradation. This scale-dependent nature is not necessarily compatible with dimensionless form such as in equation~(\ref{eq:nonlinear_law}). Moreover, it is difficult to verify the scale dependence of the physical law over a wide length scale because our laboratory scale is restricted to the meter scale. \citet{Minton:2019} suggested that ray-structured heterogeneous ejecta deposition is crucial to explaining the observed crater degradation. The physical origin of the ray structure of ejecta deposition has not yet been fully revealed. However, recent laboratory experiments have found possible ideas for crater ray structuring~\citep{Kadono:2015,Sabuwala:2018,PachecoVazquez:2019}. Hopefully, it will be possible to develop a method of accessing the scale-dependent crater degradation mechanism in the future. 

\section{Conclusion}

The migration of the centroid induced by an impact onto an inclined granular surface is analyzed using the experimental data reported in \citet{Takizawa:2020}. Based on the 3D crater profiles, centroid migration normalized to crater diameter $x_\mathrm{mig}/D_\mathrm{cy}$ is systematically measured. To explain the data behavior, we employ a nonlinear transport model (equation~(\ref{eq:nonlinear_law})) in which three parameters $\mu$, $K$, and $c$ are introduced, characterizing the effective friction coefficient of landsliding, the asymmetric ejecta-deposition effect, and the effect of oblique impact, respectively. The value of $\mu$ is close to (slightly greater than) that estimated from the angle of repose. The obtained $K$ value is consistent with the theoretical model of asymmetric ejecta deposition~\citep{Soderblom:1970}. The incident-angle dependence $c$ is  introduced in this study. However, its effect on $x_\mathrm{mig}/D_\mathrm{cy}$ is relatively minor. The experimental results extend the classical linear model~\citep{Soderblom:1970} to the nonlinear regime. Moreover, the results suggest that the linear model is sufficient for evaluating the long-term degradation of craters. The establishment of a detailed estimate of the degradation rate of crater shapes using the obtained result is an area of future study.

\section*{Acknowledgements}

We thank JSPS KAKENHI Grant No.~18H03679 for financial support. 

\section*{Data Availability}

The data will be shared on reasonable request.



\bibliographystyle{mnras}
\bibliography{ImpactRelax} 








\bsp	
\label{lastpage}
\end{document}